\documentclass[11pt]{article}

\usepackage[margin=1.1in]{geometry}
\usepackage{graphicx}

\def\Cplusplus{C\raisebox{0.35ex}{\small\textbf{++}}}

\usepackage{amssymb, amsmath, amsthm}
\usepackage{thmtools}

\begin{document}

\title{Exporting Ada Software to Python and Julia\thanks{Supported
by the National Science Foundation under grant DMS 1854513.}}

\author{Jan Verschelde\thanks{University of Illinois at Chicago,
Department of Mathematics, Statistics, and Computer Science,
851 S. Morgan St. (m/c 249), Chicago, IL 60607-7045
Email: {\tt janv@uic.edu}, URL: {\tt http://www.math.uic.edu/$\sim$jan}.}}

\date{28 June 2022}


\maketitle

\begin{abstract}
The objective is to demonstrate the making of Ada software
available to Python and Julia programmers using GPRbuild.
GPRbuild is the project manager of the GNAT toolchain.
With GPRbuild the making of shared object files is fully automated
and the software can be readily used in Python and Julia.
The application is the build process of PHCpack,
a free and open source software package to solve polynomial systems
by homotopy continuation methods, written mainly in Ada, 
with components in \Cplusplus,
available at github at 
{\tt \small https://github.com/janverschelde/PHCpack}.
\end{abstract}


\section{Language Agnostic Computing}

This paper describes interface development from the perspective of
an Ada programmer, aimed to export the functionality of a software package 
to Python~\cite{PGH11} and Julia~\cite{BEKS17} computational environments, 
available through Jupyter notebooks~\cite{Jupyter}.
The Jupyter notebook is the interface to SageMath~\cite{Ste11},
a free open source system for mathematical computing.

In order to export all functionality the interface passes through C,
which may be regarded as a least common multiple of programming languages,
as Ada, Python, and Julia share enough common ground to enable language
agnostic computing, as Jupyter stands for Julia, Python, R, and many others.

\begin{figure}[hbt]
\begin{center}
\begin{picture}(200,80)(0,0)
\put(0,60){\line(1,0){80}}  \put(0,60){\line(0,1){20}}
\put(0,80){\line(1,0){80}} \put(80,60){\line(0,1){20}}
\put(23,65){Python}
\put(40,60){\vector(1,-1){10}}
\put(100,60){\line(1,0){80}}  \put(100,60){\line(0,1){20}}
\put(100,80){\line(1,0){80}}  \put(180,60){\line(0,1){20}}
\put(128,65){Julia}
\put(140,60){\vector(-1,-1){10}}
\put(50,30){\line(1,0){80}}  \put(50,30){\line(0,1){20}}
\put(50,50){\line(1,0){80}} \put(130,30){\line(0,1){20}}
\put(63,35){C interface}
\put(90,30){\vector(0,-1){10}}
\put(50,0){\line(1,0){80}}  \put(50,0){\line(0,1){20}}
\put(50,20){\line(1,0){80}} \put(130,0){\line(0,1){20}}
\put(65,5){Ada Code}
\end{picture}
\caption{C as the least common multiple language.}
\label{figClcm}
\end{center}
\end{figure}
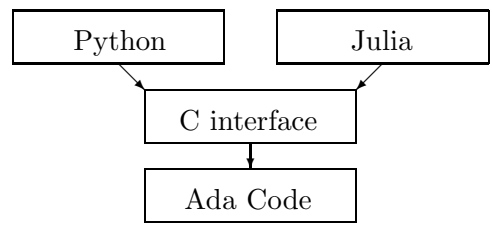

The main point is to automate building with GPRbuild.

\section{GPRbuild and Interface Development}

The mixed language development is supported by GPRbuild, the project
manager of the gnu-ada compiler GNAT.
The build process, defined via library projects, results in shared
object files (with the extension {\tt .so} on Linux, {\tt .dll} on Windows,
and {\tt .dylib} on Mac OS X).
These shared object files can be called directly from a Python script
or a Julia program.

In the C interface layer, the control is passed to a C program.
The C program passes input data to some Ada procedure, calls
an exported procedure, and extracts the output data
via another call to an Ada procedure.
The most basic and versatile manner to pass data is via a plain
sequence of characters of 32-bit integers.
As the {\em hello world} for this interface, consider the swapping
of characters in a string.

\begin{figure}[hbt]
\begin{center}
\begin{picture}(200,10)(0,8)
\put(0,8){\tt "hello"} \put(50,10){\vector(1,0){30}}
\put(80,0){\line(1,0){40}}  \put(80,0){\line(0,1){20}}
\put(80,20){\line(1,0){40}} \put(120,0){\line(0,1){20}}
\put(88,7){swap}
\put(120,10){\vector(1,0){30}} \put(155,8){\tt "olleh"}
\end{picture}
\caption{Swapping characters via an interface package.}
\label{fighello}
\end{center}
\end{figure}
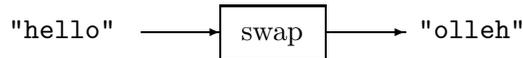

\begin{figure}[hbt]
\begin{center}
\begin{picture}(200,100)(-20,0)
\put(0,80){\line(1,0){40}}  \put(0,80){\line(0,1){20}}
\put(0,100){\line(1,0){40}} \put(40,80){\line(0,1){20}}
\put(8,87){swap}
\put(40,90){\line(1,0){100}}
\put(20,80){\line(0,-1){10}}
\put(20,65){\circle{10}} \put(25,65){\line(1,0){10}}
\put(40,62){\tt Initialize(s)}
\put(20,60){\line(0,-1){10}}
\put(20,45){\circle{10}} \put(25,45){\line(1,0){10}}
\put(40,42){\tt DoIt}
\put(20,40){\line(0,-1){10}}
\put(20,25){\circle{10}} \put(25,25){\line(1,0){10}}
\put(40,22){\tt s := Retrieve}
\put(20,20){\line(0,-1){10}}
\put(20,10){\line(1,0){120}}
\put(140,10){\line(0,1){80}}
\end{picture}
\caption{An interface package to swap characters in a string.}
\label{figinterface}
\end{center}
\end{figure}
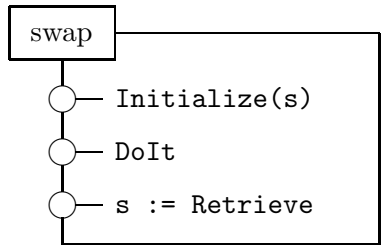

The interface package as shown in Figure~\ref{figinterface}
exports a procedure to pass the input data, the {\tt DoIt} procedure
to compute the output data, and then a third function to return
the output.

Then the C program calls the function {\tt call\_swap},
declared in Ada as below.

{\small
\begin{verbatim}
with C_Integer_Arrays; use C_Integer_Arrays;

function call_swap
  ( jobnbr : integer; sizedata : integer;
    swapdata : C_intarrs.Pointer; verbose : integer ) return integer;
\end{verbatim}
where the package {\tt C\_Integer\_Arrays}
defines {\tt C\_Integer\_Array} as
an array of C integers, of type {\tt Interfaces.C.int}.
The package contains
\begin{verbatim}
  package C_intarrs is
    new Interfaces.C.Pointers (Interfaces.C.size_T, Interfaces.C.int, C_Integer_Array,0);
\end{verbatim}
}

Observe that the {\tt void} idiom of C is avoided.
The details of this introductory project are posted at
{\tt \small github.com/janverschelde/ExportAdaGPRbuild}.

\newpage

The C code to test takes a string {\tt word},
converts the string into an array of 32-bit integers,
and then calls the Ada code:

{\small
\begin{verbatim}
sizeword = strlen(word);

for(int idx = 0; idx < sizeword; idx++)
   dataword[idx] = (int) word[idx];

adainit();
fail = _ada_call_swap(0,sizeword,dataword,1);
fail = _ada_call_swap(1,sizeword,dataword,1);
fail = _ada_call_swap(2,sizeword,dataword,1);
adafinal();

for(int idx = 0; idx < sizeword; idx++) word[idx] = (char) dataword[idx];
\end{verbatim}
}

The contents of the file {\tt demo.gpr} defines
the build of the C test program.

{\small
\begin{verbatim}
project Demo is

    for Languages use ("Ada", "C");
    for Source_Dirs use ("src");
    for Main use 
    (
        "hello_world.adb", "main.adb", "test_call_swap.c"
    );
    for Object_Dir use "obj";
    for Exec_Dir use "bin";

end Demo;
\end{verbatim}
}

To make a shared object file, a {\tt library project} is defined.
Below are the essentials of the instructions to make the
{\tt libdemo} as a shared object.

{\small
\begin{verbatim}
for Library_Dir use "lib";
for Library_Name use "demo";
for Library_Kind use "dynamic";
for Library_Auto_Init use "true";
for Library_Interface use 
(
   "hello_world", "main", "swap", "call_swap", "c_integer_arrays"
);
for Library_Standalone use "encapsulated";

package Compiler is
  for Switches ("call_swap.adb") use ("-c");
end Compiler;

package Binder is
  -- use "-Lada" for adainit and adafinal
  for Default_Switches ("Ada") use ("-n", "-Lada");
end Binder;
\end{verbatim}
}

Julia has the function {\tt ccall()} 
to execute compiled C code.
The Julia code below calls the {\tt call\_swap} procedure.

{\small
\begin{verbatim}
LIBRARY = "../Ada/lib/libdemo"

word = [Cint('h'), Cint('e'), Cint('l'), Cint('l'), Cint('o')]
println(word)
ptr2word = pointer(word, 1)
p = ccall((:_ada_call_swap, LIBRARY), Cint,
           (Cint, Cint, Ref{Cint}, Cint), 0, 5, ptr2word, 1)
p = ccall((:_ada_call_swap, LIBRARY), Cint,
           (Cint, Cint, Ref{Cint}, Cint), 1, 5, ptr2word, 1)
p = ccall((:_ada_call_swap, LIBRARY), Cint,
           (Cint, Cint, Ref{Cint}, Cint), 2, 5, ptr2word, 1)
println(word)
\end{verbatim}
}

The string {\tt "hello"} is represented by
{\tt Int32[104, 101, 108, 108, 111]}.
The last statement,
{\tt println(word)} shows {\tt Int32[111, 108, 108, 101, 104]}.

To extend Python code, an extension module must be defined in C
or \Cplusplus.  The {\tt setup.py} script has the list
{\tt extra\_objects} to define the location of the compiled Ada code
and the location of the Ada runtime libraries.
The shared object made running {\tt python setup.py build\_ext}
can then be directly imported in a Python session.
The making of this extension can be done without makefiles.

\section{Building PHCpack}

As a demonstration to a large scale project, GPRbuild is applied
to make share objects for PHCpack, a free and open source software
package to solve polynomial systems with homotopy continuation.
The python interface to PHCpack is phcpy~\cite{OFV19}.  
Written mainly in Ada,
PHCpack contains MixedVol~\cite{GLW05} and DEMiCs~\cite{MT08}
to count bounds on the number of isolated solutions fast.
For MixedVol, a translation into Ada was made.
The package DEMiCs is written in \Cplusplus\ and incorporated 
into PHCpack as such.
As described in~\cite{Ver20},
the code for multiple double precision is provided by QDlib~\cite{HLB01}
and CAMPARY~\cite{JMPT16}.

A Julia interface is under development.
From the {\tt Julia} folder of the PHCpack source distribution,
running the Julia program {\tt version.jl} at the command prompt:
{\small
\begin{verbatim}
$ julia version.jl
-> in use_c2phc4c.Handle_Jobs ...
PHCv2.4.85 released 2021-06-30
$ 
\end{verbatim}
}
The {\tt ccall()} uses the {\tt libPHCpack} shared object,
made with GPRbuild.

\bigskip

\noindent {\bf Acknowledgment.}  
The author thanks Dirk Craeynest and Fernando Oleo Blanco 
for the organization of the Ada Devroom at FOSDEM~2022.


\bibliographystyle{plain}

\end{document}